\def\Uvel{{\bf U}^{\rm flow}}
\def\Utem{{\bf U}^{\rm temp}}
\documentclass[russian,12pt]{article}
\usepackage[T1]{fontenc}
\usepackage[cp866]{inputenc}
\usepackage{babel}
\begin{document}

\begin{center}
О.М.Подвигина

~

{\bf НЕУСТОЙЧИВОСТЬ КОНВЕКТИВНЫХ ТЕЧЕНИЙ\\
МАЛОЙ АМПЛИТУДЫ ВО ВРАЩАЮЩЕМСЯ СЛОЕ\\
СО СВОБОДНЫМИ ГРАНИЦАМИ}

~

Международный институт теории прогноза землетрясений\\
и математической геофизики РАН, Москва

~

Лаборатория общей аэродинамики, Институт Механики МГУ, Москва

\end{center}

Мы изучаем устойчивость стационарных конвективных течений
в горизон\-тальном слое со свободными границами, подогреваемом снизу
и вращающемся относительно вертикальной оси, предполагая
приближение Буссинеска (конвек\-ция Рэлея-Бенара). Рассматриваемые
течения -- конвективные валы или квадрат\-ные ячейки,
являющиеся суммой двух перпендикулярных валов с одинаковыми волновыми
числами, $k$. Предполагаем, что число Рэлея близко
к критическому для возникновения конвективных течений с
волновым числом $k$: $R=R_c(k)+\varepsilon^2$;
амплитуда надкритических стационарных состояний порядка $\varepsilon$.
Показано, что течения всегда неустойчивы относительно
возмущений, являющихся суммой длинноволновой моды и двух коротковолновых
мод, соответствующих линейным валам, повернутым на малые углы
в противоположных направлениях. Макси\-мальный инкремент роста имеет порядок
$O(\max(\varepsilon^{8/5},(k-k_c)^2))$, где $k_c$ -- крити\-ческое волновое
число для установления конвекции.

\section{Введение}

Рассматривается конвекция Рэлея-Бенара в горизонтальном слое,
подогреваемом снизу и вращающемся относительно вертикальной оси.
Горизонтальные границы свободны и поддерживаются при фиксированной
температуре. В безразмерной форме система характеризуется
следующими параметрами: числами Рэлея $R$ (характеризующим
амплитуду сил плавучести), Прандтля $P$ (отношение
кинема\-тической вязкости к коэффициенту тепловой диффузии)
и Тейлора $T$ (пропорци\-ональном скорости вращения).

При малых числах Рэлея жидкость неподвижна. С увеличением числа
Рэлея в системе, в зависимости от соотношения чисел Тейлора и Прандтля,
возникает монотонная или колебательная неустойчивость \cite{chan}.
Возникающие при монотонной неустойчивости стационарные состояния
типа валов, квадратных и шестиуголь\-ных ячеек были рассмотрены, например,
в \cite{gor}-\cite{slb} (см. также монографию \cite{gez} и
обзор \cite{get}) для конвекции без вращения, и в
\cite{ver}-\cite{bass} -- для вращающегося слоя. При отсутствии
вращения все течения ответвляются в область возрастания чисел
Рэлея и, за исключением валов, неустойчивы. При
вращении возможно ветвление также в сторону убывания $R$, и
в некоторой области значений $P$ и $T$ возникающие течения с
квадратной ячейкой периодичности устойчивы по отношений
к возму\-щениям того же периода, что и у основного течения.

В работе Кюпперса и Лортца \cite{kup} было показано, что
конвективные валы могут быть неустойчивы относительно таких же валов,
повернутых на конечный угол. В пределе больших чисел Прандтля при
$T>2285$ эта неустойчивость имеет место при
установлении конвекции. Другой тип неустойчивости конвективных валов во
вращающемся слое (так называемая неустойчивость малого угла) рас\-смотрели
Кокс и Мэтьюс \cite{cox}. Используя амплитудные уравнения,
они показали, что валы с критическим волновым числом $k_c$ всегда
неустойчивы относительно возмущений, являющихся суммой длинноволновой
моды и двух экземпляров возмущаемого течения,
повернутых на малые углы порядка $\varepsilon^{2/5}$, где
$\varepsilon^2$ -- надкри\-тичность. Этот тип неустойчивости исследован
асимптотическими методами для валов, бегущих и стоящих волн для $k=k_c$
в \cite{pod}. Было показано, что существуют растущие моды при
различных углах поворота коротковолновых слагаемых.

В данной статье методы \cite{pod} применены для исследования
устойчивости ста\-ционарных течений более сложной формы с волновыми числами,
не обязательно равными критическому. Показано, что при малой надкритичности
валы и квад\-ратные ячейки, в линейном приближении
являющиеся суммой двух перпендику\-лярных валов с одинаковыми
волновыми числами $k$, неустойчивы. Максимальный инкремент роста
имеет порядок $\max(\varepsilon^{8/5},(k-k_c)^2)$.

\section{Установление конвекции во вращающемся слое при монотонной
неустойчивости}

Конвективные течения подчиняются уравнению Навье-Стокса
\begin{equation}
\label{nst}
{\partial{\bf v}\over\partial t}={\bf v}\times(\nabla\times{\bf v})
+P\Delta{\bf v}+PR\theta{\bf e}_z-\nabla p+PT{\bf v}\times{\bf e}_z
\end{equation}
с условием несжимаемости
\begin{equation}
\label{inc}
\nabla\cdot{\bf v}=0,
\end{equation}
и уравнению теплопроводности
\begin{equation}
\label{heat}
{\partial\theta\over\partial t}=-({\bf v}\cdot\nabla)\theta+v_z+\Delta\theta,
\end{equation}
где $\bf v$ -- скорость потока, $\theta$ -- разность между температурой и
ее линейным профи\-лем. На горизонтальных границах предполагается
фиксированная температура и отсутствие напряжений:
\begin{equation}
\label{bouc}
{\partial v_x\over\partial z}={\partial v_y\over\partial z}=v_z=0,
\qquad\theta=0\qquad\hbox{at }z=0,1.
\end{equation}

Система (\ref{nst})-(\ref{bouc}) допускает стационарное состояние
${\bf v}=0,\ \theta=0$. Его устойчи\-вость определяется собственными значениями
оператора линейной части системы (\ref{nst})-(\ref{bouc}). Его собственные
функции имеют вид гармоник Фурье \cite{get}. При монотонной неустойчивости
растущие моды с волновым числом $k$ существуют при $R>R_c$,
\begin{equation}
\label{rcs}
R_c(k)=(a^3+T^2\pi^2)k^{-2},\quad a=k^2+\pi^2,
\end{equation}
при колебательной -- при $R>R_c^h$
\begin{equation}
\label{rch}
R_c^h(k)=2((1+P)a^3+(PT\pi)^2(1+P)^{-1})k^{-2}.
\end{equation}
Частота колебаний удовлетворяет уравнению
\begin{equation}
\label{wcr}
\omega_c^2=T^2\pi^2a^{-1}(1-P)(1+P)^{-1}-a^2,
\end{equation}
поэтому колебательная неустойчивость возможна только при $P<1$.

Мы рассматриваем только случай монотонной
неустойчивости. Критическое волновое число $k_c$, для которого достигается
минимум $R_c(k)$, является корнем уравнения
\begin{equation}
\label{kcs}
a^2(2k^2-\pi^2)-T^2\pi^2=0.
\end{equation}

При небольшой надкритичности
\begin{equation}
\label{r2c}
R=R_c(k)+\varepsilon^2,
\end{equation}
нелинейное решение (\ref{nst})-(\ref{bouc}) можно найти,
разлагая его в ряд \cite{gez}
\begin{equation}
\label{seri}
{\bf U}=\sum_{j=1}^\infty\varepsilon^j{\bf U}_j.
\end{equation}
Здесь и ниже использовано представление полей конвективного течения
в виде\break 4-компонентных векторов, где первые 3 компоненты --
скорость потока, а последняя -- температура:
\begin{equation}
\label{Uvt}
{\bf U}_j=(\Uvel_j,\Utem_j).
\end{equation}
Первый член разложения является
собственным вектором линеаризованной сис\-темы (\ref{nst})-(\ref{bouc})
с $R=R_c(k)$, его можно представить в виде суммы
\begin{equation}
\label{sume}
{\bf U}_1=\sum_{l=1}^n {\bf F}_l,
\end{equation}
где двумерные конвективные течения ${\bf F}_l$ -- линейные валы различной
ориентации с одинаковыми волновыми числами $k$. Следующие члены
(\ref{seri}) определяются из уравнений, получающихся подстановкой (\ref{seri})
и (\ref{sume}) в (\ref{nst})-(\ref{bouc}).

Первые члены ряда \ref{seri} для $n=1$ (валы) приведены в \cite{bass}:
\begin{equation}
\label{v1}
{\bf U}_1=b \left(
\begin{array}{c}
-\pi k^{-1}\cos\pi z\sin kx\\
T\pi (ak)^{-1}\cos\pi z\sin kx\\
\sin\pi z\cos kx\\
 a^{-1}\sin\pi z\cos kx
\end{array}
\right),
\end{equation}
\begin{equation}
\label{v2}
{\bf U}_2=b^2 \left(
\begin{array}{c}
0\\
T\pi^2(8Pk^3a)^{-1}\sin 2kx\\
0\\
-(8\pi a)^{-1}\sin 2\pi z\\
\end{array}
\right),
\end{equation}
где амплитуда $b$ определяется из
\begin{equation}
\label{ampA}
8k^4P^2a=(P^2k^4R_c-T^2\pi^4)b^2.
\end{equation}
Случаи квадратных и шестиугольных ячеек рассмотрены в
\cite{ver}, \cite{gold} и \cite{gold2}.

\section{Устойчивость валов с $k=k_c$}

В этой части мы излагаем результаты работы \cite{pod}, где
исследована устойчивость валов с волновым числом, равным критическому.

Устойчивость (\ref{seri}) определяется собственными значениями оператора
$L$, явля\-ющегося линеаризацией (\ref{nst})-(\ref{heat}) вблизи этого
стационарного состояния. Этот опе\-ратор можно разложить в ряд
\begin{equation}
\label{seriL}
L=\sum_{j=0}^\infty\varepsilon^j L_j.
\end{equation}
Первые два члена имеют вид
\begin{equation}
\label{L0}
L_0({\bf v},\theta)=(P\Delta{\bf v}+PR_c\theta{\bf e}_z-\nabla p+PT{\bf v}\times{\bf e}_z,
v_z+\Delta\theta)
\end{equation}
и
\begin{equation}
\label{L1}
L_1({\bf v},\theta)=(\Uvel_1\times(\nabla\times{\bf v})+
{\bf v}\times(\nabla\times\Uvel_1),
-(\Uvel_1\cdot\nabla)\theta-({\bf v}\cdot\nabla)\Utem_1).
\end{equation}
Ниже $L_0(R)$ обозначает оператор $L_0$ при $R\ne R_c$.
Оператор $M$ определяется формулой
\begin{equation}
M({\bf v},\theta)=(P\theta{\bf e}_z,0).
\end{equation}

Для решения задачи на собственные значения
\begin{equation}
\label{eig}
L{\bf W}=\lambda\bf W
\end{equation}
вычислено инвариантное подпространство $L$; для
доказательства неустойчивости вторичного течения достаточно провести
анализ собственных значений ограни\-чения $L$ на это подпространство.

Пусть ${\bf W}_j$ ($j=1,...,J$) базис в этом подпространстве.
Обозначим через\break${\cal A}=(A_{mn})$ матрицу ограничения $L$
на это подпространство:
\begin{equation}
\label{ivs}
L{\bf W}_j=\sum_{i=1}^J A_{ij}{\bf W}_i\quad\forall\ j.
\end{equation}
Представим базисные векторы и элементы матрицы в виде рядов
\begin{equation}
\label{seriW}
{\bf W}_j=\sum_{l=0}^\infty\varepsilon^l{\bf W}_{j,l},
\end{equation}
\begin{equation}
\label{seriA}
A_{mn}=\sum_{l=0}^\infty\varepsilon^l A_{mn,l}.
\end{equation}
Подставляя (\ref{seriW}) и (\ref{seriA}) в (\ref{ivs}) и приравнивая
выражения при равных степенях $\varepsilon$, получаем систему уравнений для
${\bf W}_{j,l}$. В качестве ${\bf W}_{j,0}$ выбраны собственные векторы $L_0$:
\begin{equation}
\label{bubuW}
L_0{\bf W}_{j,0}=\lambda_{j,0}{\bf W}_{j,0}.
\end{equation}

Пусть $(\delta_x,\delta_y,0)$ -- малое возмущение волнового вектора $(k,0,\pi)$:
$\delta_x\ll k$ и $\delta_y\ll k$. Коэффициенты ${\bf W}_{j,l}$ и
$A_{mn,l}$ рядов (\ref{seriW}) и (\ref{seriA}) зависят от этих параметров.

Обозначим за $\cal F$ пространство симметричных относительно вертикальной оси
4-компонентных векторных полей вида (\ref{Uvt}), где $\Uvel$ бездивергентно,
являющих\-ся
суммами гармоник Фурье с волновыми векторами $(mk\pm\delta_x,\pm\delta_y,l\pi)$,
где $l$ и $m$ -- целые. Это пространство $L_n$-инвариантно для всех $n$.
Система уравнений для ${\bf W}_{j,l}$ имеет вид
$$L_0{\bf W}_{j,n}-\lambda_{j,0}{\bf W}_{j,n}-\sum_{i=1}^J A_{ij,n}{\bf W}_{i,0}=$$
\begin{equation}
\label{ivn}
-\sum_{m>0,l\ge0,m+l=n}L_m{\bf W}_{j,l}+\sum_{i=1}^J
\sum_{m>0,l\ge0,m+l=n}A_{ij,m}{\bf W}_{i,l}
\quad\forall\ j,
\end{equation}
где правая часть известна. Для $n=0$ решение дается (\ref{bubuW}).
Уравнения решаем последовательно для возрастающих $n\ge 1$.
Для дальнейших построений нам необходимо, чтобы решения ${\bf W}_{j,n}$
были равномерно ограничены по $\delta_x$ и $\delta_y$. Для этого достаточно,
чтобы норма ограничения $L_0^{-1}$ на инвариантное подпространс\-тво,
дополнительное в $\cal F$ к натянутому на ${\bf W}_{j,0}$, была бы
ограничена равномерно по $\delta_x$ и $\delta_y$. Поскольку у ограничения
$L_0$ на $\cal F$ существует ровно три малых собственных значения, а
остальные имеют порядок единицы, естественно, чтобы построенное
инвариантное подпространство имело размерность $J=3$.

${\bf W}_{j,0}$ имеют вид:
\begin{equation}
\label{W01}
{\bf W}_{1,0}=\left(
\begin{array}{c}
-\delta_y\sin(\delta_x x+\delta_y y)\\
\delta_x\sin(\delta_x x+\delta_y y)\\
0\\
0
\end{array}
\right),
\end{equation}
\begin{equation}
\label{W02}
{\bf W}_{2,0}=\left(
\begin{array}{c}
-(\pi s_++q_+\delta_y)k_+^{-1}
\cos\pi z\sin((k+\delta_x)x+\delta_y y)\\
(q_+s_+-\pi\delta_y k_+^{-2})
\cos\pi z\sin((k+\delta_x)x+\delta_y y)\\
\sin\pi z\cos((k+\delta_x)x+\delta_y y)\\
a_+^{-1}\sin\pi z\cos((k+\delta_x)x+\delta_y y)
\end{array}
\right)+O((k_+^2-k^2)^2),
\end{equation}
\begin{equation}
\label{W03}
{\bf W}_{3,0}=\left(
\begin{array}{c}
-(\pi s_--q_-\delta_y)k_-^{-1}
\cos\pi z\sin((k-\delta_x)x-\delta_y y)\\
(q_-s_-+\pi\delta_y k_-^{-2})
\cos\pi z\sin((k-\delta_x)x-\delta_y y)\\
\sin\pi z\cos((k-\delta_x)x-\delta_y y)\\
a_-^{-1}\sin\pi z\cos((k-\delta_x)x-\delta_y y)
\end{array}
\right)+O((k_-^2-k^2)^2).
\end{equation}
Здесь $k_\pm=((k\pm\delta_x)^2+\delta_y^2)^{1/2}$,
$s_\pm=(k\pm\delta_x)k_\pm^{-1}$, $a_\pm=k_\pm^2+\pi^2$ и
$q_\pm=T\pi(k_\pm a_\pm)^{-1}$. Собственные векторы (\ref{W01})-(\ref{W03})
имеют собственные значения
\begin{equation}
\label{lambda10}
\lambda_{1,0}=-P(\delta_x^2+\delta_y^2),
\end{equation}
$$\lambda_{2,0}=-C_1Pg^{-1}a^{-1}(k_+^2-k^2)^2+O((k_+^2-k^2)^3),$$
$$\lambda_{3,0}=-C_1Pg^{-1}a^{-1}(k_-^2-k^2)^2+O((k_-^2-k^2)^3),$$
где $C_1={1\over 2}\partial^2 R_c/\partial(k^2)^2=3(\pi^2+k^2)/k^2$ и
$g={1\over 4}((2\pi^2-k^2)k^{-2}+3P)$.

Элементы матрицы $\cal A$ имеют вид
\begin{equation}
\label{AA}
\begin{array}{lll}
A_{11}&=&-P(\delta_x^2+\delta_y^2)+O(\varepsilon^2\delta^2),\\
A_{21}&=&
{1\over 2}kb\varepsilon\delta_y+\varepsilon F_1(\delta^2)+O(\varepsilon\delta^3,\varepsilon^3),\\
A_{31}&=&
-{1\over 2}kb\varepsilon\delta_y+\varepsilon F_1(\delta^2)+O(\varepsilon\delta^3,\varepsilon^3),\\
A_{12}&=&
\varepsilon(\delta_x^2+\delta_y^2)^{-1}
(C_2{\pi b\over{2k}}(\delta_y^2-\delta_x^2)+{b\over 2}\delta_x\delta_y(C_2^2-{{\pi^2}\over{k^2}})
+\\
&&F_2(\delta^3))+O(\varepsilon\delta^2,\varepsilon^3)),\\
A_{22}&=&-\varepsilon^2C_3-C_1Pg^{-1}a^{-1}(k_+^2-k^2)^2+\\
&&O((k_+^2-k^2)^3,\varepsilon^2\delta,\varepsilon^4),\\
A_{32}&=&-\varepsilon^2C_3+O(\varepsilon^2\delta,\varepsilon^4),\\
A_{13}&=&
\varepsilon(\delta_x^2-\delta_y^2)^{-1}
(C_2{\pi b\over{2k}}(\delta_y^2-\delta_x^2)+{b\over 2}\delta_x\delta_y(C_2^2-{{\pi^2}\over{k^2}})-\\
&&-F_2(\delta^3))+O(\varepsilon\delta^2,\varepsilon^3)),\\
A_{23}&=&-\varepsilon^2C_3+O(\varepsilon^2\delta,\varepsilon^4),\\
A_{33}&=&-\varepsilon^2C_3-C_1Pg^{-1}a^{-1}(k_-^2-k^2)^2+\\
&&O((k_-^2-k^2)^3,\varepsilon^2\delta,\varepsilon^4),
\end{array}
\end{equation}
где
$C_2=T\pi(ka)^{-1}$, $F_1=f_1\delta_x^2+f_2\delta_x\delta_y+f_3\delta_y^2$,
$F_2=f_4\delta_x^3+f_5\delta_x^2\delta_y+f_6\delta_x\delta_y^2+f_7\delta_y^3$,
а величины $C_3$ и $f_i$ несущественны.

Рассматривая матрицу $\cal A$ в новом базисе
\begin{equation}
\label{babazis}
\widetilde{\bf W}_1={\bf W}_1,
\widetilde{\bf W}_2={\bf W}_2+{\bf W}_3,
\widetilde{\bf W}_3={\bf W}_2-{\bf W}_3
\end{equation}
получаем \cite{pod}, что при $\delta_x^2+\delta_y^2\gg\varepsilon^2$
собственные значения матрицы (\ref{AA}) асимпто\-тически близки к
$-P(\delta_x^2+\delta_y^2)$ и собственным значениям матрицы
\begin{equation}
\label{AA0}
\left[
\begin{array}{cc}
\widetilde A_{22}-\xi_2\widetilde A_{12}&\widetilde A_{23}-\xi_2\widetilde A_{13}\\
\widetilde A_{32}-\xi_3\widetilde A_{12}&\widetilde A_{22}-\xi_3\widetilde A_{13}
\end{array}
\right],
\end{equation}
где $\xi_2=\widetilde A_{21}/(\widetilde A_{11}-\widetilde A_{22})$ и
$\xi_3=\widetilde A_{31}/(\widetilde A_{11}-\widetilde A_{33})$
($\widetilde{\cal A}$ -- это матрица $\cal A$, записанная в новом базисе
(\ref{babazis})~).

Собственные значения (\ref{AA0}) асимптотически равны
\begin{equation}
\label{lam0}
\begin{array}{l}
\lambda=-C_4(4k^2\delta_x^2+\delta_y^4)-C_3\varepsilon^2\pm
[2C_4k\delta_x\delta_y^2(8C_4k\delta_x\delta_y^2+
b^2\varepsilon^2\delta_y(\delta_x^2+\delta_y^2)^{-1}\times\\
(C_2\pi(\delta_y^2-\delta_x^2)+k\delta_x\delta_y(C_2^2-{{\pi^2}\over{k^2}}))
(P(\delta_x^2+\delta_y^2)+C_4\delta_x^2)^{-1})+C_3\varepsilon^2)]^{1/2},
\end{array}
\end{equation}
где $C_4=C_1Pg^{-1}a^{-1}$.
Максимальное по $\delta_x$ и $\delta_y$ собственное значение
\begin{equation}
\label{lamm}
\lambda_{\max}={5\over 4}Pg^{-1}k^{-2}\delta_y^4
\end{equation}
достигается при
\begin{equation}
\label{xymm}
3k^2\delta_x^2=\delta_y^4\hbox{  и  }
\delta_y^5=-{\sqrt 3\over 12}P^{-2}T\pi^2a^{-1}gb^2\varepsilon^2k.
\end{equation}

При $\delta_x^2+\delta_y^2=O(\varepsilon^2)$ собственные значения
(\ref{AA}) по порядку не превосходят $O(\varepsilon^2)$, что меньше
$\lambda_{\max}\sim\varepsilon^{8/5}$.

Оператор, сопряженный к $L_0$ относительно скалярного произведения
\begin{equation}
\label{scpro}
({\bf w}_1,{\bf w}_2)={\bf w}_1^{\rm flow}\cdot{\bf w}_2^{\rm flow}
+PR_c{\bf w}_1^{\rm temp}\cdot{\bf w}_2^{\rm temp},
\end{equation}
совпадает с $L_0$, в котором направление вращения изменено на противоположное
(здесь $\cdot$ обозначает обычное скалярное произведение в пространстве
${\bf L}_2$). Базис, двойственный к ${\bf W}_{j,0}$, имеет вид
\begin{equation}
\label{dual}
\begin{array}{l}
{\bf W}_{1,0}^*=(\delta_x^2+\delta_y^2)^{-1}{\bf W}_{1,0},\\
{\bf W}_{2,0}^*=g_+^{-1}{\bf W}_{2,0}(-T)+O((k_+^2-k^2)^2)\hbox{  и}\\
{\bf W}_{3,0}^*=g_-^{-1}{\bf W}_{3,0}(-T)+O((k_-^2-k^2)^2),$$
\end{array}
\end{equation}
где
$$g_+=({\bf W}_{2,0},{\bf W}_{2,0}(-T)),\qquad
g_-=({\bf W}_{3,0},{\bf W}_{3,0}(-T)).$$

\section{Устойчивость валов с $k\ne k_c$}

Анализ устойчивости валов при $k\ne k_c$ проводим аналогично
исследованию для валов с волновым числом, равным критическому. Рассматриваем
собственные значения оператора $L$, являющегося линеаризацией
(\ref{nst})-(\ref{heat}) вблизи стационарного состояния (\ref{seri}),
(\ref{v1}), (\ref{v2}). Оператор записан в виде (\ref{seriL}),
первые члены разложения задаются выражениями (\ref{L0}) и (\ref{L1}).

Трехмерное инвариантное пространство $L$ с базисом ${\bf W}_j$, $j=1,2,3$,
строится так же, как и для $k=k_c$. Элементы матрицы
$\cal A$ ограничения $L$ на это подпростран\-ство и базисные векторы
представлены в виде рядов (\ref{seriA}) и (\ref{seriW}),
где ${\bf W}_{j,0}$ -- собственные векторы $L_0$.

Собственное значение $\lambda_{1,0}$ (\ref{lambda10}) не зависит от $k$;
$\lambda_{2,0}$ и $\lambda_{3,0}$ имеют вид \cite{pod}
\begin{equation}
\label{bubula}
\begin{array}{l}
\lambda_{2,0}=(R_c(k)-R_+)(M{\bf W}_{j,0},{\bf W}_{j,0}^*)+O((R_c(k)-R_+)^2),\\
\lambda_{3,0}=(R_c(k)-R_-)(M{\bf W}_{j,0},{\bf W}_{j,0}^*)+O((R_c(k)-R_-)^2),
\end{array}
\end{equation}
где $R_\pm=R_c(k_\pm)$ -- критическое число Рэлея для $k_\pm$ (\ref{rcs}).
Разложим его в ряд Тейлора в окрестности $k$:
$$R_\pm=R_c(k)+\partial R_c/\partial(k^2)(k_\pm^2-k^2)+
{1\over 2}\partial^2 R_c/\partial(k^2)^2(k_\pm^2-k^2)^2+$$
\begin{equation}
\label{tabu}
O((k_\pm^2-k^2)^3)=
R_c+\partial^2 R_c/\partial^2(k^2)(k^2-k_c^2)(k_\pm^2-k^2)+
\end{equation}
$${1\over 2}\partial^2 R_c/\partial(k^2)^2(k_\pm^2-k^2)^2+
O((k^2-k_c^2)(k_\pm^2-k^2)^2,(k_\pm^2-k^2)^3).$$
При $k=k_c$ выполнено $\partial R_c(k)/\partial(k^2)=0$.

При $k\ne k_c$ член порядка $\varepsilon$ ($l=1$) ряда (\ref{seriA})
для матрицы (\ref{AA}) не изменяется, т.к. при его вычислении в \cite{pod}
условие $k=k_c$ не было использовано.
Таким образом, в силу (\ref{bubula}), (\ref{tabu}) и (\ref{AA}), для
$k\ne k_c$ главные члены матрицы $\cal A$ ограничения оператора $L$ на
трехмерное инвариантное пространство теперь имеют вид:

\begin{equation}
\label{AAl}
\begin{array}{lll}
A_{11}&=&-P(\delta_x^2+\delta_y^2)+O(\varepsilon^2\delta^2),\\
A_{21}&=&
{1\over 2}kb\varepsilon\delta_y+\varepsilon F_1(\delta^2)+O(\varepsilon\delta^3,\varepsilon^3),\\
A_{31}&=&
-{1\over 2}kb\varepsilon\delta_y+\varepsilon F_1(\delta^2)+O(\varepsilon\delta^3,\varepsilon^3),\\

A_{12}&=&
\varepsilon(\delta_x^2+\delta_y^2)^{-1}
(C_2{\pi b\over{2k}}(\delta_y^2-\delta_x^2)+{b\over 2}\delta_x\delta_y(C_2^2-{{\pi^2}\over{k^2}})+\\
&&F_2(\delta^3))+O(\varepsilon\delta^2,\varepsilon^3)),\\
A_{22}&=&-\varepsilon^2C_3-C_1Pg^{-1}a^{-1}(k_+^2-k^2)^2
-4C_1Pg^{-1}a^{-1}\alpha k(k_+^2-k^2)+\\
&&O(\alpha(k_+^2-k^2)^2,(k_+^2-k^2)^3,\varepsilon^2\delta,\varepsilon^4),\\
A_{32}&=&-\varepsilon^2C_3+O(\varepsilon^2\delta,\varepsilon^4),\\

A_{13}&=&
\varepsilon(\delta_x^2-\delta_y^2)^{-1}
(C_2{\pi b\over{2k}}(\delta_y^2-\delta_x^2)+{b\over 2}\delta_x\delta_y(C_2^2-{{\pi^2}\over{k^2}})-\\
&&F_2(\delta^3))+O(\varepsilon\delta^2,\varepsilon^3)),\\
A_{23}&=&-\varepsilon^2C_3+O(\varepsilon^2\delta,\varepsilon^4),\\
A_{33}&=&-\varepsilon^2C_3-C_1Pg^{-1}a^{-1}(k_-^2-k^2)^2
-4C_1Pg^{-1}a^{-1}\alpha k(k_-^2-k^2)+\\
&&O(\alpha(k_-^2-k^2)^2,(k_-^2-k^2)^3,\varepsilon^2\delta,\varepsilon^4),
\end{array}
\end{equation}
где $\alpha=k-k_c$.

При $\delta_x^2+\delta_y^2\gg\varepsilon^2$, рассматривая матрицу
$\cal A$ в новом базисе (\ref{babazis}) как и при $k=k_c$, получаем,
что наибольшее собственное значение (\ref{AAl}) асимптотически равно
$$\lambda=-C_4(4k^2\delta_x^2+\delta_y^4+4\alpha k\delta_y^2)-C_3\varepsilon^2+$$
\begin{equation}
\label{laml}
[(2C_4k\delta_x\delta_y^2+4\alpha k^2\delta_x)
8C_4k\delta_x\delta_y^2+16\alpha k^2\delta_x+
b^2\varepsilon^2\delta_y(\delta_x^2+\delta_y^2)^{-1}\times
\end{equation}
$$(C_2\pi(\delta_y^2-\delta_x^2)+k\delta_x\delta_y(C_2^2-{{\pi^2}\over{k^2}}))
(P(\delta_x^2+\delta_y^2)+C_4\delta_x^2)^{-1})+C_3\varepsilon^2)]^{1/2}.$$
В зависимости от соотношения $\alpha$ и $\varepsilon$, максимум
(\ref{laml}) достигается при различных условиях на $\delta_x$ и $\delta_y$.

\bigskip
Случай 1: $\alpha\ll\varepsilon^{4/5}.$

Наибольшее собственное значение асимптотически задается выражением
(\ref{lamm}), достигаемого при условиях (\ref{xymm}).
В этом случае дополнительные члены матрицы (\ref{AAl}) порядка
$\sim (k^2-k_c^2)(k_\pm^2-k^2)$ несущественны,
и неустойчивость типа малого угла является доминирующей.

Случай 2: $\alpha\gg\varepsilon^{4/5}.$

Максимальное собственное значение асимптотически равно
\begin{equation}
\label{lam2}
\lambda=4C_4\alpha^2k^2
\end{equation}
и достигается при
\begin{equation}
\label{lam22}
2k\delta_x+\delta_y^2=-2\alpha k.
\end{equation}
Это неустойчивость типа Экхауза\footnote{Неустойчивость Экхауза соответствует
случаю $\delta_y=0$. Как хорошо известно (см., например, \cite{get,eck}, а также
ссылки в \cite{tb}), неустойчивость Экхауза -- это неустойчивость валов с
волновым вектором $(k,0,\pi)$ относительно валов с волновым вектором
$(q,0,\pi)$, имеющая место для чисел Рэлея, удовлетворяющих неравенству
$3(R-R_c(k))\le R_c(k)-R_c(k_c)$, где $R_c(k)$ определено (\ref{rcs}).}.
Заметим, что (\ref{lam22}) позволяет определить только волновое число наиболее
неустойчивой моды, для определения направления волнового вектора
при максимизации $\lambda$ в нем необходимо учитывать членов следующего
порядка малости по всем малым параметрам.

Случай 3: $\alpha\sim\varepsilon^{4/5}$.

Этот случай наиболее интересен и сложен --
имеет место взаимодействие двух неустойчивостей.
Наибольший по $\delta_x$ и $\delta_y$ инкремент
роста имеет порядок $\sim\varepsilon^{8/5}$, и
$\delta_x\sim\varepsilon^{4/5}$ и $\delta_y\sim\varepsilon^{2/5}$.
Мы не приводим здесь точные выражения $\lambda_{\max}$ и компонент
критического волнового вектора,
поскольку они чересчур громоздки и неинформативны.

\section{Устойчивость квадратных ячеек}

Поскольку горизонтальный слой обладает вращательной инвариантностью,
воз\-можны более сложные стационарные течения, чем рассмотренные
выше валы (см., например, \cite{get}).
Для таких течений первый член ряда (\ref{seri}) имеет вид (\ref{sume}):
\begin{equation}
\label{qq}
{\bf U}^{\rm sq}_1=\sum_{i=1}^n b_i\gamma_{\varphi_i}{\bf U}_1,
\end{equation}
где ${\bf U}_1$ -- линейные валы (\ref{seri}), $\gamma_{\varphi_i}$
-- оператор поворота на угол $\varphi_i$ вокруг вертикаль\-ной оси.
Амплитуды могут быть определены, например, методом, предложенным
Веронисом \cite{ver} (см. также \cite{gez}). Мы рассматриваем только случай
квадратных ячеек, т.е.
$n=2$, $\varphi_1=0$ и $\varphi_2=\pi/2$. Течения такой геометрии с
$b_1=b_2$ при установлении конвекции устойчивы относительно короткопериодных
возмущений в некоторой области значений $P$ и $T$ \cite{gold}.

Для исследования устойчивости квадратных ячеек построим инвариантное
пространство оператора $L^{\rm sq}$, являющегося линеаризацией (\ref{nst})-(\ref{heat})
в окрестности этого стационарного состояния, размерности пять. В качестве
приближений нуле\-вого порядка по $\varepsilon$ базисных полей этого
инвариантного пространства выбираем
$${\bf W}_{1,0},{\bf W}_{2,0},{\bf W}_{3,0},$$
\begin{equation}
\label{bus}
{\bf W}_{4,0}(\delta_x,\delta_y)=\gamma_{\pi/2}{\bf W}_{2,0}(-\delta_y,\delta_x),
\end{equation}
$${\bf W}_{5,0}(\delta_x,\delta_y)=
\gamma_{\pi/2}{\bf W}_{3,0}(-\delta_y,\delta_x).$$
Двойственный базис определяется аналогично:
$${\bf W}_{1,0}^*,{\bf W}_{2,0}^*,{\bf W}_{3,0}^*,$$
$${\bf W}_{4,0}^*(\delta_x,\delta_y)=
\gamma_{\pi/2}{\bf W}_{2,0}^*(-\delta_y,\delta_x),$$
$${\bf W}_{5,0}^*(\delta_x,\delta_y)=
\gamma_{\pi/2}{\bf W}_{3,0}^*(-\delta_y,\delta_x).$$

Член разложения (\ref{seriL}) нулевого порядка не зависит от потока, следующий
член имеет вид
$$L^{\rm sq}_1=L_1({\bf U}_1)+L_1(\gamma_{\pi/2}{\bf U}_1)$$
(здесь $L_1({\bf U})$ обозначает оператор (\ref{L1}), вычисленный для
стационарного состояния $\bf U$). Элементы матрицы $\cal A$, определяющей
устойчивость квадратных ячеек, можно определить, зная элементы соответствующей
матрицы (\ref{AAl}) для валов.

Рассмотрим вначале случай $k=k_c$. Тогда
\begin{equation}
\label{AAsq}
\begin{array}{lll}
A_{11}&=&-P(\delta_x^2+\delta_y^2)+O(\varepsilon^2\delta^2),\\
A_{21}&=&
{1\over 2}kb_1\varepsilon\delta_y+\varepsilon F_1(\delta^2)+O(\varepsilon\delta^3,\varepsilon^3),\\
A_{31}&=&
-{1\over 2}kb_1\varepsilon\delta_y+\varepsilon F_1(\delta^2)+O(\varepsilon\delta^3,\varepsilon^3),\\
A_{41}&=&
{1\over 2}kb_2\varepsilon\delta_x+\varepsilon F_1(\delta^2)+O(\varepsilon\delta^3,\varepsilon^3),\\
A_{51}&=&
-{1\over 2}kb_2\varepsilon\delta_x+\varepsilon F_1(\delta^2)+O(\varepsilon\delta^3,\varepsilon^3),\\

A_{12}&=&
\varepsilon(\delta_x^2+\delta_y^2)^{-1}
(C_2{\pi b_1\over{2k}}(\delta_y^2-\delta_x^2)+{b_1\over 2}\delta_x\delta_y(C_2^2-{{\pi^2}\over{k^2}})+\\
&&F_2(\delta^3))+O(\varepsilon\delta^2,\varepsilon^3)),\\
A_{22}&=&-\varepsilon^2C_3-C_1Pg^{-1}a^{-1}(k_+^2-k^2)^2+
O((k_+^2-k^2)^3,\varepsilon^2\delta,\varepsilon^4),\\
A_{32}&=&-\varepsilon^2C_3+O(\varepsilon^2\delta,\varepsilon^4),\\

A_{13}&=&
\varepsilon(\delta_x^2-\delta_y^2)^{-1}
(C_2{\pi b_1\over{2k}}(\delta_y^2-\delta_x^2)+{b_2\over 2}\delta_x\delta_y(C_2^2-{{\pi^2}\over{k^2}})-\\
&&F_2(\delta^3))+O(\varepsilon\delta^2,\varepsilon^3)),\\
A_{23}&=&-\varepsilon^2C_3+O(\varepsilon^2\delta,\varepsilon^4),\\
A_{33}&=&-\varepsilon^2C_3-C_1Pg^{-1}a^{-1}(k_-^2-k^2)^2+O((k_-^2-k^2)^3,\varepsilon^2\delta,\varepsilon^4),\\

A_{14}&=&
\varepsilon(\delta_x^2+\delta_y^2)^{-1}
(C_2{\pi b_2\over{2k}}(\delta_x^2-\delta_y^2)+{b_2\over 2}\delta_x\delta_y(C_2^2-{{\pi^2}\over{k^2}})+\\
&&F_4(\delta^3))+O(\varepsilon\delta^2,\varepsilon^3)),\\
A_{44}&=&-\varepsilon^2C_5-C_1Pg^{-1}a^{-1}(\tilde k_+^2-k^2)^2+
O((\tilde k_+^2-k^2)^3,\varepsilon^2\delta,\varepsilon^4),\\
A_{54}&=&-\varepsilon^2C_5+O(\varepsilon^2\delta,\varepsilon^4),\\

A_{15}&=&
\varepsilon(\delta_x^2+\delta_y^2)^{-1}
(C_2{\pi b_2\over{2k}}(\delta_x^2-\delta_y^2)+{b_2\over 2}\delta_x\delta_y(C_2^2-{{\pi^2}\over{k^2}})-\\
&&F_4(\delta^3))+O(\varepsilon\delta^2,\varepsilon^3)),\\
A_{45}&=&-\varepsilon^2C_5+O(\varepsilon^2\delta,\varepsilon^4),\\
A_{55}&=&-\varepsilon^2C_5-C_1Pg^{-1}a^{-1}(\tilde k_-^2-k^2)^2+
O((\tilde k_-^2-k^2)^3,\varepsilon^2\delta,\varepsilon^4)
\end{array}
\end{equation}
где $\tilde k_\pm=((k\pm\delta_y)^2+\delta_x^2)^{1/2}$; $g$, $C_1$, $F_1$, $F_2$
те же, что и для валов, а значения $C_3$, $C_5$ и $f_i$ несущественны.
Коэффициенты, не приведенные в (\ref{AAsq}), имеют порядок $O(\varepsilon^3)$.

Пусть $\delta_y\gg\delta_x$. Из асимптотики коэффициентов
(\ref{AAsq}) следует, что, при\break$\delta_x^2+\delta_y^2\gg\varepsilon^2$,
$A_{44}$ и $A_{55}$ асимптотически больше остальных коэффициентов
$A_{4i}$, $A_{i4}$, $A_{5i}$ и $A_{i5}$ $\forall i$. Следовательно,
у матрицы (\ref{AAsq}) есть отрицательные
собственные значения, асимптотически близкие к $A_{44}$ и $A_{55}$,
которым отвечают собственные векторы
$$\tilde{\bf W}_4={\bf W}_4+\xi_1 {\bf W}_1\quad\hbox{и}$$
$$\tilde{\bf W}_5={\bf W}_5+\xi_2 {\bf W}_1,$$
где $\xi_1=A_{14}/(A_{11}+A_{44})$ и $\xi_2=A_{15}/(A_{11}+A_{55})$.
Представим матрицу в новом базисе ${\bf W}_4\to\widetilde{\bf W}_4$,
${\bf W}_5\to\widetilde{\bf W}_5$. Оставшиеся три собственные значения матрицы
(\ref{AAsq}) тогда равны собственным значениям ее подматрицы размером
$3\times3$ в верхнем левом углу, которая имеет вид:
\begin{equation}
\label{AAmod}
\begin{array}{lll}
\widetilde A_{ij}&=&A_{ij}\hbox{ для }(ij)\ne (21),(31);\\
\widetilde A_{21}&=&A_{21}-\xi_1 A_{12};\\
\widetilde A_{31}&=&A_{31}-\xi_2 A_{12},
\end{array}
\end{equation}
где $\widetilde A_{ij}$, $i,j=1,2,3$, -- элементы матрицы (\ref{AAsq}),
записанной в базисе $\widetilde{\bf W}$.
Дополни\-тельные члены, возникающие при изменении базиса, асимптотически
малы, и матрицы (\ref{AA}) и (\ref{AAmod}) совпадают (с точностью до замены
$b$ на $b_1$). Максимальное собственное значение (\ref{AA}), найденное
в разделе (3), задается формулой (\ref{lamm}).

Для $k\ne k_c$, так же как и при исследовании устойчивости конвективных
валов, в матрице (\ref{AAsq}) на диагонали появляются дополнительные
слагаемые:\break$4C_1Pg^{-1}a^{-1}(k_+^2-k^2)$, $4C_1Pg^{-1}a^{-1}(k_-^2-k^2)$,
$4C_1Pg^{-1}a^{-1}(\tilde k_+^2-k^2)$ и $4C_1Pg^{-1}a^{-1}(\tilde k_-^2-k^2)$
в элементах $A_{22}$, $A_{33}$, $A_{44}$ и $A_{55}$, соответственно.
Рассматривая три варианта соотношений $(k_c-k)$ и $\varepsilon$, так же,
как в конце раздела (4), и используя описанную выше замену базиса
${\bf W}_4\to\widetilde{\bf W}_4$, ${\bf W}_5\to\widetilde{\bf W}_5$,
можно показать, что для каждого из трех вариантов соотношений
существует растущая мода. При этом, как и в случае валов, доминируют три
типа неустойчивости: при $\alpha\ll\varepsilon^{4/5}$
неустойчивость типа малого угла, при $\alpha\gg\varepsilon^{4/5}$ --
типа Экхауза, а при $\alpha\sim\varepsilon^{4/5}$ неустойчивости
этих двух типов взаимодействуют.

\section{Заключение}

Мы показали, что рассмотренные конвективные течения (валы и квадратные ячейки)
во вращающемся слое всегда неустойчивы относительно длинноволновых возмущений.
В зависимости от соотношений надкритичности и разности волново\-го числа течения и
критического волнового числа, доминирует либо неустойчи\-вость
малого угла, либо неустойчивость Экхауза. Вычисления приведены для случая
надкритического ветвления стационарных конвективных состояний, одна\-ко если они
ответвляются в область уменьшения $R$ (см. \cite{gold}), результаты сохраня\-ются.

Длинноволновая неустойчивость конвективных течений доминирует над
не\-устойчивостью по отношению к коротковолновым возмущениям (если таковая
имеет место), поскольку у нее инкремент роста асимптотически больше
($O(\varepsilon^{8/5})$ для неустойчивости малого угла, когда
волновое число основного течения близко ко критическому, или
$O((k-k_c)^2)\gg O(\varepsilon^{8/5})$ для неустойчивости типа Экхауза,
имеющей место в противном случае) асимптотически больше инкремента роста
для коротковолновой неустойчивости, имеющего порядок $O(\varepsilon^2)$.

Для пространственных структур, у которых первый член разложения по
$\varepsilon$ является суммой трех или более
валов вида (\ref{qq}), можно провести анализ устойчивости аналогичным образом,
однако, с увеличением числа валов размер исследуемой на собственные значения
матрицы возрастает и вычисления стано\-вятся более громоздкими.

\bigskip
Работа выполнена при поддержке Российского фонда фундаментальных
ис\-следований (грант 04-05-64699).

\pagebreak


\begin{thebibliography}{99}
\bibitem{chan}
{\it Chandrasekhar S.} Hydrodynamic and hydromagnetic stability. Oxford:
Claredon press, 1961, 652 c.

\bibitem{gor}
{\it Горьков Л.П.} Стационарная конвекция в плоском слое жидкости вблизи
критического режима теплопередачи// ЖЭТФ. 1957. Т.~33. С.~402--407.

\bibitem{mal}
{\it Malkus W.V.R., Veronis G.} Finite amplitude cellular convection //
J.~Fluid Mech. 1959. V.~4. P.~225--260.

\bibitem{slb}
{\it Schluter A., Lortz D., Busse F.H.} On the stability of steady finite
amplitude convection // J.~Fluid Mech. 1965 V.~23. P.~129--144.

\bibitem{gez}
{\it Гершуни Г.З., Жуховицкий Е.М.} Конвективная устойчивость несжимаемой
жидкости. М.: Наука, 1972, 392 с.

\bibitem{get}
{\it Гетлинг А.И.} Формирование пространственных структур конвекции
Рэлея-Бенара // УФН. 1991. Т.~161 С.~1--80.

\bibitem{ver}
{\it Veronis G.} Cellular convection with finite amplitude in a rotating
fluid // J.~Fluid Mech. 1959. V.~5. P.~401--435.

\bibitem{kup}
{\it Kuppers G., Lortz D.} Transition from laminar convection to thermal
turbulence in a rotating fluid layer // J.~Fluid Mech. 1969. V.~35. P.~609--620.

\bibitem{gold}
{\it Goldstein H.F., Knobloch E., Silber M.} Planform selection in rotating
convection // Phys. Fluids A. 1990. V.~2. P.~625--627.

\bibitem{gold2}
{\it Goldstein H.F., Knobloch E., Silber M.} Planform selection in rotating
convection: Hexagonal symmetry // Phys Rev. A. 1992. V.~46. P.~4755--4761.

\bibitem{bass}
{\it Bassom P.B., Zhang K.} Strongly nonlinear convection cells in a rapidly
rotating fluid layer // Geophys. Astrophys. Fluid Dynamics. 1994. V.~76.
P.~223--238.

\bibitem{cox}
{\it Cox S.M., Matthews P.C.} Instability of rotating convection //
J.~Fluid Mech. 2000. V.~403. P.~153--172.

\bibitem{pod}
{\it Podvigina O.M.} Instability of flows near the onset of convection
in a rotating layer with stress-free horizontal boundaries //
подано в J.~Fluid Mech.

\bibitem{eck}
{\it Eckhaus W.} Studies in Nonlinear Stability Theory. Berlin: Springer, 1965.

\bibitem{tb}
{\it Tuckerman L.S., Barkley D.} Bifurcation analysis of the Eckhaus
instability // Physica D 1990. V.~46. P.~57--86.
\end{thebibliography}
\end{document}